\documentstyle[12pt]{article}
\def\simgt{\stackrel{>}{{}_\sim}}
\def\be{\begin{equation}}
\def\ee{\end{equation}}
\def\bear{\be\begin{array}}
\def\eear{\end{array}\ee}
\def\bea{\begin{eqnarray}}
\def\eea{\end{eqnarray}}

\def\baselinestretch{1}
\begin{document}
\catcode`@=11
\newtoks\@stequation
\def\subequations{\refstepcounter{equation}%
\edef\@savedequation{\the\c@equation}%
  \@stequation=\expandafter{\theequation}
  \edef\@savedtheequation{\the\@stequation}
  \edef\oldtheequation{\theequation}%
  \setcounter{equation}{0}%
  \def\theequation{\oldtheequation\alph{equation}}}
\def\endsubequations{\setcounter{equation}{\@savedequation}%
  \@stequation=\expandafter{\@savedtheequation}%
  \edef\theequation{\the\@stequation}\global\@ignoretrue

\noindent}
\catcode`@=12
\begin{titlepage}
\title{{\bf Stability Bounds on Flavor-Violating Trilinear Soft Terms
in the MSSM}
\thanks{Research supported in part by: the CICYT, under
contract AEN95-0195 (JAC); the European
Union, under contract CHRX-CT92-0004 (JAC).}
}
\author{ {\bf J.A. Casas\thanks{On leave of absence from Instituto de
Estructura de la Materia CSIC, Serrano 123, 28006 Madrid, Spain.}
${}^{ {\footnotesize,\S}}$},
and {\bf S. Dimopoulos${}^{\footnotesize\P}$}\\
\hspace{3cm}\\
${}^{\footnotesize\S}$ {\small Santa Cruz Institute for Particle
Physics}\\
{\small University of California, Santa Cruz, CA 95064, USA}\\
\vspace{-0.3cm}\\
${}^{\footnotesize\P}$ {\small Theoretical Physics Division, CERN}\\
{\small CH--1211 Geneva 23, Switzerland}\\
{\small and}\\
{\small Physics Department, Stanford University}\\
{\small Stanford CA 94305, USA}}
\date{}
\maketitle
\def\baselinestretch{1.15}

\begin{abstract}
\noindent
The stability of the standard vacuum imposes constraints on flavor
violating trilinear soft terms which are stronger than the laboratory
bounds coming from the absence of neutral flavor violations (FCNC).
Furthermore, contrary to the FCNC bounds, these constraints persist
even if the scale of supersymmetry breaking is arbitrarily large.

\end{abstract}

\thispagestyle{empty}

\vspace{3cm}
{\small
\leftline{CERN-TH-96/116}
\leftline{SCIPP-96-22}
\leftline{SU-ITP 96/16}
\leftline{IEM-FT-131/96}
\leftline{May 1996}}
\vskip-22.5cm
{\small
\rightline{CERN-TH-96/116}
\rightline{SCIPP-96-22}
\rightline{SU-ITP 96/16}
\rightline{IEM-FT-131/96}}
\rightline{hep-ph/9606237}

\end{titlepage}
\newpage
\setcounter{page}{1}

\section{Introduction}

Supersymmetry has sources of flavor violation which are not
present in the Standard Model \cite{Dim81}. These arise from the
possible
presence of non-diagonal terms in the squark and slepton mass
matrices,
which come from the soft breaking Lagrangian, ${\cal L}_{\rm soft}$:
\footnote{We work
in a basis for the superfields where the Yukawa coupling matrices
are diagonal.}
\bea
\label{nodiag}
-{\cal L}_{\rm soft}&=&
\left(m_L^2\right)_{ij} \bar L_{L_i}L_{L_j}\ +\
\left(m_{e_R}^2\right)_{ij} \bar{e}_{R_i} e_{R_j}\
\nonumber \\
&+&\
\left(m_Q^2\right)_{ij} \bar Q_{L_i}Q_{L_j}\ +\
\left(m_{u_R}^2\right)_{ij} \bar{u}_{R_i} u_{R_j}\ +\
\left(m_{d_R}^2\right)_{ij} \bar{d}_{R_i} d_{R_j}
\nonumber \\
&+&\  \left[ A^l_{ij}\bar L_{L_i} H_1 e_{R_j} + A^u_{ij}
\bar Q_{L_i} H_2 u_{R_j}
+ A^d_{ij}Q_{L_i} H_1 d_{R_j}
+ {\rm h.c.} \right]\; +\ \cdots \; ,
\eea
where $i,j=1,2,3$ are generation indices. A usual simplifying
assumption
of the MSSM is that $m^2_{ij}$ is  diagonal and universal and
$A_{ij}$
is proportional to the corresponding Yukawa matrix. However, there is
no compelling theoretical argument for these hypotheses.

The size of the off-diagonal entries in $m^2_{ij}$, $A_{ij}$
is strongly restricted by FCNC experimental data\cite{Dim81, Gabbi89,
Hagel94,
Choud95, Barbi94, Dim95, Car95, Gabbi96}. In this paper we will
focus our attention on the $A_{ij}^{(f)}$ terms; a summary of the
corresponding FCNC bounds is given in the second column of
Table 1 \cite{Choud95, Gabbi96}. The
$\left(\delta^{(f)}_{LR}\right)_{ij}$ parameters used in the table
are
defined as
\bea
\label{notation}
\left(\delta^{(f)}_{LR}\right)_{ij}\equiv
\frac{\left(\Delta
M_{LR}^{2\;\;(f)}\right)_{ij}}{M^{2\;\;(f)}_{av}}\;\;,
\eea
where $f=u,d,l$;
$M^{2\;\;(f)}_{av}$ is the average of the squared sfermion
($\tilde f_L$ and $\tilde f_R$) masses
and $\left(\Delta M_{LR}^{2\;\;(f)}\right)_{ij}$ are the off-diagonal
entries in the sfermion mass matrices
\bea
\label{MLR}
\left(\Delta M_{LR}^{2\;\;(u)}\right)_{ij}=A^{(u)}_{ij}
\langle H_2^0\rangle, \hspace{0.4cm}
\left(\Delta M_{LR}^{2\;\;(d)}\right)_{ij}=A^{(d)}_{ij}
\langle H_1^0\rangle\;,\hspace{0.4cm}
\left(\Delta M_{LR}^{2\;\;(l)}\right)_{ij}=A^{(l)}_{ij}\langle H_1^0
\rangle\;.
\eea

In this paper we show that the $A_{ij}^{(f)}$ terms are
also restricted on completely different grounds, namely from the
requirement
of the absence of dangerous charge and color breaking (CCB) minima or
unbounded from below (UFB) directions. As we will see, these bounds
are
in general {\em stronger than the FCNC ones}. Other properties of
these bounds are the following:

\begin{description}

\item[{\em i)}] Some of the bounds, particularly the UFB ones, are
genuine
effects of the non-diagonal $A_{ij}^{(f)}$ structure, i.e. they do
not
have a ``diagonal counterpart''.

\item[{\em ii)}] Contrary to the FCNC bounds, the strength of the
CCB and UFB bounds does not decrease as the scale of supersymmetry
breaking increases.

\end{description}

In sections 2 and 3 we derive the bounds. In section 4 we discuss
their implication for various theories.

\section{Constraints on $A_{ij}$ from CCB and UFB}

Let us start with a CCB bound.
Consider the off-diagonal trilinear scalar coupling
\be
\label{emu}
A^{(l)}_{12}\ \bar{e}_L H_1^0 \mu_R\ + \ {\rm h.c.}
\ee
Along the field-space direction $|e_L|=|H_1^0|=|\mu_R|\equiv a$,
the $SU(3)\times SU(2)\times U(1)$ D-terms are vanishing. Since
the phases of the fields can be chosen so that (\ref{emu}) becomes
negative, the relevant terms of the potential are
\bea
\label{Vccb}
V&=&m_{e_L}^2 |e_L|^2 + m_{\mu_R}^2 |\mu_R|^2 + m_{1}^2 |H_1|^2
+  |\lambda_e \bar{e}_L H_1^0|^2
+  |\lambda_\mu H_1^0 \mu_R|^2
 - 2\left|A^{(l)}_{12}\right| \bar{e}_L H_1^0 \mu_R
\nonumber\\
&=&\left(m_{e_L}^2 + m_{\mu_R}^2 + m_1^2\right)a^2\ +\
\left(\lambda_e^2 + \lambda_\mu^2\right)a^4\ -\
2\left|A^{(l)}_{12}\right|a^3\ \ .
\eea
Neglecting the $\lambda_e^2$ term, it is
straightforward to check that a deep CCB minimum appears
at $a\sim 2|A_{12}^{(l)}|/\lambda_{\mu}^2$ unless
\be
\label{ccbbound}
\left|A^{(l)}_{12}\right|^2 \leq \lambda_\mu^2\left(m_{e_L}^2 +
m_{\mu_R}^2 + m_1^2\right)
\ee
This bound is analogous to the ``traditional'' CCB bounds
\cite{Frere}
for {\em diagonal}
$A$--terms, namely $\left|A^{(l)}_{11}\right|^2 \leq 3\lambda_e^2
\left(m_{e_L}^2 + m_{e_R}^2 + m_1^2\right)$,
$\left|A^{(l)}_{22}\right|^2
\leq 3\lambda_\mu^2 \left(m_{\mu_L}^2 + m_{\mu_R}^2 + m_1^2\right)$.
The bound (\ref{ccbbound}) is easily generalized to other couplings
\bea
\label{ccbbound2}
\left|A^{(u)}_{ij}\right|^2 &\leq& \lambda_{u_k}^2\left(m_{u_{L_i}}^2
+
m_{u_{R_j}}^2 + m_2^2\right),
\hspace{1cm}k={\rm Max}\ (i,j)
\nonumber\\
\left|A^{(d)}_{ij}\right|^2 &\leq& \lambda_{d_k}^2\left(m_{d_{L_i}}^2
+
m_{d_{R_j}}^2 + m_1^2\right),
\hspace{1cm}k={\rm Max}\ (i,j)
\nonumber\\
\left|A^{(l)}_{ij}\right|^2 &\leq& \lambda_{e_k}^2\left(m_{e_{L_i}}^2
+
m_{e_{R_j}}^2 + m_1^2\right),
\hspace{1cm}k={\rm Max}\ (i,j)
\eea
These bounds are, in general, stronger than
the corresponding FCNC ones. Actually, they
can be made more restrictive by considering extra scalar
fields in the potential. This works in the same way as for the
``traditional'' CCB bounds \cite{CCB1,CCB2}.
For example, for the first two generations the
right hand side of eqs.(\ref{ccbbound2}) can be modified as
$m_1^2\rightarrow m_1^2-\mu^2$, $m_2^2\rightarrow m_2^2-\mu^2$
(for more details see ref.\cite{CCB2}). Other possible improvements
are more model-dependent, but do not change the order
of magnitude of the bounds.

\vspace{0.25cm}
\noindent
Let us now derive a simple UFB bound. Consider again the trilinear
term
of eq.(\ref{emu}) and the following direction in the (scalar)
field--space
\be
\label{ufbdir}
|e_L|^2=|\mu_R|^2 =|\nu_\tau|^2 + |H_1^0|^2\equiv a^2\;,
\ee
which of course requires $|H_1|^2<a^2$. Then, the $SU(2)\times U(1)$
D--terms get cancelled,
\bea
\label{VD}
V_{D-terms}&=&\frac{1}{8}g_2^2\left[|H_1^0|^2 + |\nu_\tau|^2
- |e_L|^2 \right]^2\nonumber\\
&+& \frac{1}{8}{g'}^2\left[|H_1^0|^2 + |\nu_\tau|^2
+ |e_L|^2 - 2|\mu_R|^2 \right]^2 \ =\ 0\;,
\eea
so that the scalar potential is given by
\bea
\label{Vufb}
V&=&a^2\left[m_{e_L}^2\ +\ m_{\mu_R}^2\ +\ m_{\nu_\tau}^2\
-\ 2|A^{(l)}_{12}|\ H_1^0\ +\  |\lambda_\mu H_1^0|^2
\right]\nonumber\\
&+&\left(m_1^2- m_{\nu_\tau}^2\right) |H_1^0|^2\;.
\eea
Minimizing with respect to $H_1$, we find
\be
\label{H11}
H_1^0= \frac{|A^{(l)}_{12}|a^2}{\lambda_\mu^2a^2 +
\left(m_1^2- m_{\nu_\tau}^2\right)}\;,
\ee
which satisfies $|H_1|^2<a^2$ for large enough values of $a$. Then,
the potential of eq.(\ref{Vufb}) becomes
\bea
\label{Vufb1}
V=a^2\left[m_{e_L}^2\ +\ m_{\mu_R}^2\ +\ m_{\nu_\tau}^2\
-\ |A^{(l)}_{12}|^2 \frac{a^2}{\lambda_\mu^2a^2+
\left(m_1^2- m_{\nu_\tau}^2\right)}\right]\;.
\eea
So, the potential becomes deeply negative unless
\be
\label{ufbbound1}
\left|A^{(l)}_{12}\right|^2\frac{a^2}{\lambda_\mu^2a^2+
\left(m_1^2- m_{\nu_\tau}^2\right)}
 \leq \left(m_{e_L}^2 +
m_{\mu_R}^2 + m_{\nu_\tau}^2\right)
\ee
The above equation should be satisfied for any value of $a$ such that
$|H_1^0|$, as given by (\ref{H11}), satisfies $|H_1|^2<a^2$.
An interesting limit occurs for
$a\gg\frac{\left(m_1^2- m_{\nu_\tau}^2\right)}{\lambda_\mu^2}$.
Then
\be
\label{H1}
H_1^0= \frac{|A^{(l)}_{12}|}{\lambda_\mu^2}\;,
\ee
and, provided  $a^2>|H_1|^2$, the potential (\ref{Vufb1}) reads
\bea
\label{Vufb2}
V=a^2\left[m_{e_L}^2\ +\ m_{\mu_R}^2\ +\ m_{\nu_\tau}^2\
-\ \frac{|A^{(l)}_{12}|^2}{\lambda_\mu^2}
\right]
\eea
and the previous bound (\ref{ufbbound1}) simply becomes
\be
\label{ufbbound}
\left|A^{(l)}_{12}\right|^2 \leq \lambda_\mu^2\left(m_{e_L}^2 +
m_{\mu_R}^2 + m_{\nu_\tau}^2\right)
\ee

The previous UFB example is useful to check the property {\em (i)}
mentioned in the introduction. Indeed, it is easy to
verify that this kind of UFB direction cannot take place
if $A_{ij}$ is diagonal (property {\em (i)})
since in that instance the quartic part of the $H_1$ F-term would
not be vanishing. Also, notice that property {\em (ii)}
is a consequence of the fact that both members of eq.(\ref{ufbbound})
scale in the same way as the typical supersymmetry breaking mass
increases. Of course, this also holds for the CCB constraints
summarized in eqs. (\ref{ccbbound2}).

Let us now extend this UFB constraint to the other trilinear terms.
For the leptonic ones eq.(\ref{ufbbound})
is inmediately generalized to
\bea
\label{ufbbound2}
\hspace{0.2cm}
\left|A^{(l)}_{ij}\right|^2 \leq \lambda_{e_k}^2\left(m_{e_{L_i}}^2 +
m_{e_{R_j}}^2 + m_{\nu_m}^2\right),
\hspace{0.8cm}k={\rm Max}\ (i,j),\;\; m\neq i,j.
\eea
For the $A^{(d)}_{ij}$ terms things work in a very similar way,
interchanging $e\leftrightarrow d$. More precisely, taking
the following direction in the (scalar) field--space
\be
\label{ufbdir3}
|d_{L_i}|^2=|d_{R_j}|^2 =|\nu_m|^2 + |H_1^0|^2\equiv a^2\;,\;\; m\neq
i,j
\ee
the $SU(3)\times SU(2)\times U(1)$ D--terms get cancelled,
\bea
\label{VD2}
V_{D-terms}&=&\frac{1}{6}g_3^2\left[|d_{L_i}|^2 -|d_{R_j}|^2
\right]^2\nonumber\\
&+&\frac{1}{8}g_2^2\left[|H_1^0|^2 + |\nu_m|^2
- |d_{L_i}|^2 \right]^2\nonumber\\
&+& \frac{1}{8}{g'}^2\left[|H_1^0|^2 + |\nu_m|^2
-\frac{1}{3} |d_{L_i}|^2 -\frac{2}{3}|d_{R_j}|^2 \right]^2 \ =\ 0\;.
\eea
Therefore, following the same steps as in the leptonic case, see
eqs. (\ref{Vufb})--(\ref{ufbbound2}), we arrive at the corresponding
UFB
bound for $A^{(d)}_{ij}$ terms
\bea
\label{ufbbound3}
\left|A^{(d)}_{ij}\right|^2 \leq \lambda_{d_k}^2\left(m_{d_{L_i}}^2 +
m_{d_{R_j}}^2 + m_{\nu_m}^2\right),
\hspace{1cm}k={\rm Max}\ (i,j)
\eea

For the $A^{(u)}_{ij}$ things change since we need the contribution
of two sleptons of different generations $|e_{L_p}|^2,|e_{R_q}|^2$
($p\neq q$), rather than $|\nu_{m}|^2$, in order to cancel the
D--terms.
To see this, notice that in this case the D--terms have the form
(we take $m\neq p,q$ for simplicity)
\bea
\label{VD3}
V_{D-terms}&=&\frac{1}{6}g_3^2\left[|u_{L_i}|^2 -|u_{R_j}|^2
\right]^2\nonumber\\
&+&\frac{1}{8}g_2^2\left[|H_2^0|^2 - |\nu_m|^2
+|e_{L_p}|^2  - |u_{L_i}|^2 \right]^2\nonumber\\
&+& \frac{1}{8}{g'}^2\left[|H_2^0|^2 - |\nu_m|^2
-|e_{L_p}|^2 + 2|e_{R_q}|^2 +\frac{1}{3} |u_{L_i}|^2
-\frac{4}{3}|u_{R_j}|^2 \right]^2 \;,
\eea
which are cacelled for
\bea
\label{ufbdir4}
|\nu_m|^2&=&0,\hspace{0.5cm}
|e_{L_p}|^2=|e_{R_q}|^2,\nonumber\\
|u_{L_i}|^2&=&|u_{R_j}|^2 =|e_{L_p}|^2 + |H_2^0|^2\equiv a^2\;.
\eea
So, following again the steps of eqs.
(\ref{Vufb})--(\ref{ufbbound2}),
we obtain the corresponding UFB bound for $A^{(u)}_{ij}$ terms
\bea
\label{ufbbound4}
\hspace{0.2cm}
\left|A^{(u)}_{ij}\right|^2 \leq \lambda_{u_k}^2\left(m_{u_{L_i}}^2 +
m_{u_{R_j}}^2 + m_{e_{L_p}}^2 + m_{e_{R_q}}^2\right),
\hspace{0.8cm}k={\rm Max}\ (i,j),\;\; p\neq q\;.
\eea
The UFB bounds can also be slightly improved by
considering extra scalar fields. However, the simplified
limits of the bounds, i.e. eqs. (\ref{ufbbound2},
\ref{ufbbound3}, \ref{ufbbound4}), cannot be modified in a simple
model-independent way.

\vspace{0.25cm}
\noindent
The CCB and UFB bounds collected in eqs.(\ref{ccbbound2},
\ref{ufbbound2}, \ref{ufbbound3}, \ref{ufbbound4}) must be imposed
at a renormalization scale, $Q$, of the order of the VEVs of the
relevant fields. This means that the CCB bounds must be evaluated
at a scale $Q\sim 2A_{ij}^{(f)}/\lambda_{f_k}^2$, while the
UFB bounds must be imposed at any possible value of $Q\sim a$.
This can be relevant in many instances. For example,
for universal gaugino
and scalar masses ($M_{1/2}$ and $m$ respectively) satisfying
$M_{1/2}\simgt m$, the UFB bounds are more restrictive at $M_X$ than
at low energies (especially the hadronic ones). This trend gets
stronger
as the ratio $M_{1/2}/m$ increases.

\section{Numerical results}

Let us express the previously obtained CCB and UFB bounds in terms
of the $\left(\delta^{(f)}_{LR}\right)_{ij}$ parameters defined in
eqs.(\ref{notation}, \ref{MLR}). The CCB bounds,
eqs.(\ref{ccbbound2}),
read
\bea
\label{Deltabound1}
\left(\delta^{(l)}_{LR}\right)_{ij} &\leq&
{\cal M}_{e_k}
\frac{\left[2M^{2\;\;(l)}_{av}+m_1^2\right]^{1/2}}{M^{2\;\;(l)}_{av}}
\hspace{1cm}k={\rm Max}\ (i,j)
\nonumber\\
\left(\delta^{(d)}_{LR}\right)_{ij} &\leq&
{\cal M}_{d_k}
\frac{\left[2M^{2\;\;(d)}_{av}+m_1^2\right]^{1/2}}{M^{2\;\;(d)}_{av}}
\hspace{1cm}k={\rm Max}\ (i,j)
\nonumber\\
\left(\delta^{(u)}_{LR}\right)_{ij} &\leq&
{\cal M}_{u_k}
\frac{\left[2M^{2\;\;(u)}_{av}+m_2^2\right]^{1/2}}{M^{2\;\;(u)}_{av}}
\hspace{1cm}k={\rm Max}\ (i,j),
\eea
while the UFB bounds,
eqs.(\ref{ufbbound2}, \ref{ufbbound3}, \ref{ufbbound4}), can be
essentially expressed as
\bea
\label{Deltabound2}
\left(\delta^{(l)}_{LR}\right)_{ij} &\leq&
{\cal M}_{e_k}\frac{\sqrt{3}}{M^{(l)}_{av}}
\hspace{2cm}k={\rm Max}\ (i,j)
\nonumber\\
\left(\delta^{(d)}_{LR}\right)_{ij} &\leq&
{\cal M}_{d_k}
\frac{\left[2M^{2\;\;(d)}_{av}+M^{2\;\;(l)}_{av}\right]^{1/2}}
{M^{2\;\;(d)}_{av}}
\hspace{1cm}k={\rm Max}\ (i,j)
\nonumber\\
\left(\delta^{(u)}_{LR}\right)_{ij} &\leq&
{\cal M}_{u_k}
\frac{\left[2M^{2\;\;(u)}_{av}+2M^{2\;\;(l)}_{av}\right]^{1/2}}
{M^{2\;\;(u)}_{av}}
\hspace{1cm}k={\rm Max}\ (i,j).
\eea
In eqs.(\ref{Deltabound1}, \ref{Deltabound2})
${\cal M}_{f_k}$ represents
the mass of the fermion $f_k$.

These CCB and UFB bounds are almost always stronger than the
corresponding FCNC bounds. This is
illustrated in Table 1 for the particular case ${\cal M}_{f_k}=500$
GeV.
The only exception is $\left(\delta^{(l)}_{LR}\right)_{12}$, which is
experimentally constrained by the $\mu\rightarrow e,\gamma$ process.
As the scale of supersymmetry breaking increases the FCNC bounds are
easily
satisfied whereas the CCB and UFB bounds continue to strongly
constrain the theory.
Another case in which the FCNC constraints are satisfied
is when approximate ``infrared universality'' emerges from the RG
equations
\cite{gaugino, Choud95, Car95}. Again, the CCB and UFB bounds
continue to impose
strong constraints on such theories. This is because, as argued
before, these bounds have to be evaluated at different
large scales and do not benefit from RG running.

\begin{table}[h]

\begin{center}
\begin{tabular}{||c|c|c||}\hline
\hspace{0.5cm}&\hspace{0.5cm}&\hspace{0.5cm}\\
\hspace{0.5cm}& FCNC & CCB and UFB \\
\hspace{0.5cm}&\hspace{0.5cm}&\hspace{0.5cm}\\ \hline
\hspace{0.5cm}&\hspace{0.5cm}&\hspace{0.5cm}\\
$\left(\delta^{(d)}_{LR}\right)_{12}$ & $4.4 \times 10^{-3}$ &
$2.9\times 10^{-4}$\\
\hspace{0.5cm}&\hspace{0.5cm}&\hspace{0.5cm}\\
$\left(\delta^{(d)}_{LR}\right)_{13}$ &  $3.3 \times 10^{-2}$&
$10^{-2}$\\
\hspace{0.5cm}&\hspace{0.5cm}&\hspace{0.5cm}\\
$\left(\delta^{(d)}_{LR}\right)_{23}$ &  $1.6\times 10^{-2}$&
$10^{-2}$\\
\hspace{0.5cm}&\hspace{0.5cm}&\hspace{0.5cm}\\
$\left(\delta^{(u)}_{LR}\right)_{12}$ &  $3.1\times 10^{-2}$&
$2.3\times 10^{-3}$\\
\hspace{0.5cm}&\hspace{0.5cm}&\hspace{0.5cm}\\
$\left(\delta^{(l)}_{LR}\right)_{12}$ &  $8.5\times 10^{-6}$&
$3.6\times 10^{-4}$\\
\hspace{0.5cm}&\hspace{0.5cm}&\hspace{0.5cm}\\
$\left(\delta^{(l)}_{LR}\right)_{13}$ &  $5.5\times 10^{-1}$&
$6.1\times 10^{-3}$\\
\hspace{0.5cm}&\hspace{0.5cm}&\hspace{0.5cm}\\
$\left(\delta^{(l)}_{LR}\right)_{23}$ &  $10^{-1}$&
$6.1\times 10^{-3}$ \\
\hspace{0.5cm}&\hspace{0.5cm}&\hspace{0.5cm}\\
\hline
\end{tabular}
\end{center}

\caption{FCNC bounds versus  CCB and UFB  bounds on
$\left(\delta^{(f)}_{LR}\right)_{ij}$ for
$M_{av}^{(f)}=500$ GeV.
The bounds have been obtained from ref.[8]
taking $x=(m_{\rm gaugino}/M_{av}^{(f)})^2=1$.}

\end{table}

\section{Implications for supersymmetric and superstring models}

\subsection{Fritzsch models}

In the Fritzsch ansatz \cite{Fritz},
the  Yukawa--coupling
matrices of quarks in the interaction basis have the following
texture
\bea
\label{Fritzsch}
\lambda^{(u)}=\left(
\begin{array}{ccc}
0&\sqrt{\lambda_u\lambda_c}&0\\
\sqrt{\lambda_u\lambda_c}&0&\sqrt{\lambda_c\lambda_t}\\
0&\sqrt{\lambda_c\lambda_t}&\lambda_t
\end{array}
\right)\;,\;\;
\lambda^{(d)}=\left(
\begin{array}{ccc}
0&\sqrt{\lambda_d\lambda_s}&0\\
\sqrt{\lambda_d\lambda_s}&0&\sqrt{\lambda_s\lambda_b}\\
0&\sqrt{\lambda_s\lambda_b}&\lambda_b
\end{array}
\right)\;,
\eea
where the magnitude of the entries is to be understood in an
approximate
way. The (1,3), (3,1) entries can be filled, if desired, following
the same pattern (e.g.
$\lambda^{(d)}_{13}\sim\sqrt{\lambda_d\lambda_b}$)
with almost no effect in the results. By analogy, a similar texture
can be assumed for the lepton matrix, $\lambda^{(l)}_{ij}$.

Let us make the further assumption that the trilinear soft terms,
$A_{IJK}\phi_I\phi_J\phi_K$, are such that
\bea
\label{prop}
A_{IJK}\sim O(1)\times M_{SUSY}\times\lambda_{IJK}\;,
\eea
where $\lambda_{IJK}$ is the corresponding Yukawa coupling in the
superpotential. This
occurs in simple SUGRA scenarios.

Then the $A^{(f)}_{ij}$ matrices, in the basis where
the fermion matrices are diagonalized,  have essentially the same texture
as the Fritzsch matrices, i.e.
\bea
\label{FritzschA}
A^{(f)}_{ij}\sim O(1)\times M_{SUSY}\times \lambda^{(f)}_{ij}\;.
\eea
Now it is easy to check that the CCB and UFB conditions obtained
in the previous section (see e.g.
eqs.(\ref{Deltabound1}, \ref{Deltabound2})) are in general
automatically fulfilled since typically $|A^{(f)}_{ij}|\propto
\sqrt{\lambda_i\lambda_j}<\lambda_k$ with $k={\rm Max}\ (i,j)$.

These models, therefore, are safe with respect to CCB and UFB
bounds. This, however, is the exception rather than the rule and in
general the CCB and UFB
bounds strongly constraint theories.
Consider, for example, the so called Democratic scenarios
\cite{democratic}
in which all the elements of the
fermion mass matrices are very close to 1. In these theories the
approximate proportionality of
equation 27 is inadequate and the CCB and UFB constraints impose
severe constraints on the models.

\subsection{Superstring Scenarios}

The most interesting application of the CCB and UFB constraints
obtained here is to generic SUGRA frameworks, particularly
superstring scenarios.

The general SUGRA expression for $A_{IJK}$,
 as defined in eq.(\ref{nodiag}),
is given by \cite{Hall83, Kaplu93}
\bea
\label{AIJL}
A_{IJL}=\frac{1}{3}F^\phi\left[
\partial_\phi \lambda_{IJL} - \Gamma^N_{\phi\left(I\right.}
\lambda_{\left.JL\right)N}+\frac{1}{2}(\partial_\phi K)\lambda_{IJL}
\right]\;,\;\;
\eea
with
\bea
\label{AIJL2}
F^\phi=
e^{K/2}g^{\phi\bar\phi'}\left(\partial_{\phi'} W
+ ({\partial_{\phi'}K)W} \right),\; g_{\bar IJ}\equiv
\partial^2 K/\partial \Phi_I\partial\Phi_J,\;
\Gamma^N_{\phi I}\equiv g^{N\bar J}\partial_\phi g_{\bar J I} .
\eea
$K$ and $W$ are the K\"ahler potential and the original SUGRA
superpotential respectively,
while $I,J,L,N$, as well as $\phi$ run over all the chiral fields.
 $\lambda_{IJL}$ are the Yukawa couplings in the effective
superpotential (they are a factor $e^{K/2}$ the original ones).

It is clear from (\ref{AIJL}) that only the third term in the right
hand
side satisfies the proportionality relation (\ref{prop}).
The second term is particularly relevant
for our concern
since it mixes different Yukawa couplings through the non-trivial
structure
of the K\"ahler manifold. Thus, the constraints summarized in
eqs.(\ref{Deltabound1}, \ref{Deltabound2}) put non-trivial
constraints on geometric properties
of the SUGRA structure.

In superstring theories the K\"ahler metric, $g_{IJ}$, for the
observable
fields depends at tree level on the moduli, $T_i$, and at higher
orders
also on the dilaton, $S$ \cite{Kaplder}.
The Yukawa couplings depend at tree level on the moduli,
with some exceptions, as for the untwisted fields in orbifold
constructions.
As a consequence, the general situation is $A_{IJL}=O(m_{3/2})$
\cite{Kaplu93, Louis95}.
A exception to this rule occurs for dilaton-dominance SUSY breaking,
i.e. when only $F_S\neq 0$. In that case, working at tree level,
one gets exact universality of the soft breaking terms (although this
universality is spoiled at higher orders). However, for
moduli-dominated SUSY breaking this will {\em not} be the general
case.

Here, rather than making a detailed analysis of all the
possibilities,
we will illustrate the relevance of the CCB and UFB bounds by
applying
them to an interesting superstring inspired model that has been
discussed
in the literature \cite{Choud95}. Namely, consider the following
ansatz for
the $A^{(f)}$ matrices
\bea
\label{Choudhury}
A^{(f)}=\left(
\begin{array}{ccc}
0&0&A^f\\
0&0&A^f\\
A^f&A^f&A^f
\end{array}
\right)\;,\;\;
\eea
where $f=u,d,l$. The off-diagonal entries in this matrix are
useful in order to decrease the low--energy magnitude of the
additional
sources of flavor violation $\left(\Delta m^{(f)}_{LL}\right)^2$,
$\left(\Delta m^{(f)}_{RR}\right)^2$, through the RG
running provided $A^f$ is large enough.

The authors of ref.[4] showed that for $A^f=O(1)\times m_{3/2}$ the
model is safe with respect to {\em all} the FCNC constraints, even if
the
initial values of $\left(\Delta m^{(f)}_{LL,RR}\right)^2$ are
$O(m_{3/2})$,
provided that a moderate hierarchy $M_{1/2}/m_{3/2}=O(10)$ takes
place.
A slighter, but still appreciable, improvement is obtained by setting
$A^u=O(\lambda_t)\times m_{3/2}$, $A^d=O(\lambda_b)\times m_{3/2}$,
$A^l=O(\lambda_\tau)\times m_{3/2}$. It is interesting to stress
that, from the previous discussion on  $A$--terms in superstring
constructions, this scenario may occur in the framework
of superstrings if the $A$-terms associated to the lighter
generations
are small.

Nevertheless, it is easy to check that this model does not survive
the CCB and UFB bounds (see e.g.
eqs.(\ref{Deltabound1}, \ref{Deltabound2}). Even in the mentioned
(less strong) case in which $A^f$ is proportional to the largest
Yukawa coupling of the $f$-type, the CCB and UFB bounds can only be
satisfied by proper chice of the
values of the various O(1) constants.

\section{ Remarks}

The CCB and UFB bounds presented here are conservative;
they correspond to sufficient, but not necessary, conditions for the
stability of the
standard vacuum. It is possible
that we live in a metastable vacuum \cite{Claudson, Kusenko},
whose lifetime is
longer than the age of the universe. This softens
the constraints obtained here. However, it is conceptually difficult
to understand how  the cosmological constant is vanishing precisely
in a local  ``interim'' vacuum. It is also interesting that 
many of the UFB
directions found here are really unbounded from below and, if present,
 make the theory ill defined until Planckean physics comes to the
rescue.

To conclude, the stability bounds presented here are one more
manifestation of the supersymmetric flavor problem
\cite{Dim95,Car95}. They have the unique feature that they cannot be
satisfied by simply increasing the scale of supersymmetry breaking.
The simplest cure to all the
flavor problems is found in theories where supersymmetry breaking
originates at
low energies and is communicated to the ordinary sparticles via gauge
interactions.

%
%



\def\MPL #1 #2 #3 {{\em Mod.~Phys.~Lett.}~{\bf#1}\ (#2) #3 }
\def\NPB #1 #2 #3 {{\em Nucl.~Phys.}~{\bf B#1}\ (#2) #3 }
\def\PLB #1 #2 #3 {{\em Phys.~Lett.}~{\bf B#1}\ (#2) #3 }
\def\PR  #1 #2 #3 {{\em Phys.~Rep.}~{\bf#1}\ (#2) #3 }
\def\PRD #1 #2 #3 {{\em Phys.~Rev.}~{\bf D#1}\ (#2) #3 }
\def\PRL #1 #2 #3 {{\em Phys.~Rev.~Lett.}~{\bf#1}\ (#2) #3 }
\def\PTP #1 #2 #3 {{\em Prog.~Theor.~Phys.}~{\bf#1}\ (#2) #3 }
\def\RMP #1 #2 #3 {{\em Rev.~Mod.~Phys.}~{\bf#1}\ (#2) #3 }
\def\ZPC #1 #2 #3 {{\em Z.~Phys.}~{\bf C#1}\ (#2) #3 }

\end{document}